\documentclass[%
 twocolumn,
superscriptaddress,
 amsmath,amssymb,
 aps,
pra,
]{revtex4-1}

\usepackage{graphicx}
\usepackage{dcolumn}
\usepackage{bm}
\usepackage{tikz}
\usepackage[colorlinks=true,urlcolor=blue,citecolor=blue,linkcolor=blue]{hyperref}
\usepackage{orcidlink}
\usepackage{comment}
\usetikzlibrary{quantikz}
\usepackage{physics}
\usepackage{amsfonts}
\usepackage{dsfont}
\usepackage{algorithm}
\usepackage{algorithmic}
\usepackage{braket}
\usepackage{ragged2e}
\usepackage{amsmath}
\usepackage{mathtools}
\usepackage[mathlines]{lineno}
\usepackage[normalem]{ulem}
\usepackage{amsthm}
\usepackage{multirow}
\usepackage{bm}
\usepackage{dsfont}
\usepackage{makecell}
\usepackage{amssymb}
\usepackage{physics}
\usepackage{mathtools}
\usepackage{xcolor}\usepackage{soul}
\usepackage{etoolbox}
\usepackage{graphicx}
\usepackage{mwe}

\DeclareMathOperator{\EE}{\mathbb{E}}
\newcommand\E[1]{\ensuremath\EE[#1]}

\newcommand{\HS}{\mathcal{H}}
\newcommand{\K}{{\mathcal{K}}}
\newcommand{\F}{\mathcal{F}}
\newcommand{\w}{\omega}
\newcommand{\herm}{{\textrm{Herm}(\HS)}}
\DeclareMathOperator{\spann}{span}
\newcommand\spanset[1]{\ensuremath\spann(#1)}

\newcommand{\dop}{{d_\textrm{op}}}

\newcommand{\inprod}[2]{\langle #1, #2 \rangle}

\theoremstyle{definition}

\theoremstyle{remark}

\theoremstyle{remark}

\makeatletter
\DeclareFontFamily{OMX}{MnSymbolE}{}
\DeclareSymbolFont{MnLargeSymbols}{OMX}{MnSymbolE}{m}{n}
\SetSymbolFont{MnLargeSymbols}{bold}{OMX}{MnSymbolE}{b}{n}
\DeclareFontShape{OMX}{MnSymbolE}{m}{n}{
    <-6>  MnSymbolE5
   <6-7>  MnSymbolE6
   <7-8>  MnSymbolE7
   <8-9>  MnSymbolE8
   <9-10> MnSymbolE9
  <10-12> MnSymbolE10
  <12->   MnSymbolE12
}{}
\DeclareFontShape{OMX}{MnSymbolE}{b}{n}{
    <-6>  MnSymbolE-Bold5
   <6-7>  MnSymbolE-Bold6
   <7-8>  MnSymbolE-Bold7
   <8-9>  MnSymbolE-Bold8
   <9-10> MnSymbolE-Bold9
  <10-12> MnSymbolE-Bold10
  <12->   MnSymbolE-Bold12
}{}

\let\llangle\@undefined
\let\rrangle\@undefined
\DeclareMathDelimiter{\llangle}{\mathopen}%
                     {MnLargeSymbols}{'164}{MnLargeSymbols}{'164}
\DeclareMathDelimiter{\rrangle}{\mathclose}%
                     {MnLargeSymbols}{'171}{MnLargeSymbols}{'171}
\makeatother

\DeclarePairedDelimiter{\bbra}{\llangle}{\rvert}
\DeclarePairedDelimiter{\kket}{\lvert}{\rrangle}
\newcommand{\bbrakket}[2]{ \llangle{#1}\lvert{#2}\rrangle}

\begin{document}

\title{Dual frame optimization for informationally complete quantum measurements}

\author{Laurin E. Fischer \orcidlink{0000-0002-4557-8418}}\altaffiliation{These authors contributed equally to this work}
\affiliation{IBM Quantum, IBM Research Europe -- Zurich, 8803 R\"{u}schlikon, Switzerland}
\affiliation{Theory and Simulation of Materials (THEOS), {\'E}cole Polytechnique F{\'e}d{\'e}rale de Lausanne, 1015 Lausanne, Switzerland}
\author{Timoth\'ee Dao \orcidlink{0009-0009-2344-3529}}\altaffiliation{These authors contributed equally to this work}
\affiliation{IBM Quantum, IBM Research Europe -- Zurich, 8803 R\"{u}schlikon, Switzerland}
\affiliation{Institute for Theoretical Physics, ETH Z\"{u}rich, 8093 Z\"{u}rich, Switzerland}
\author{Ivano Tavernelli \orcidlink{0000-0001-5690-1981}}
\affiliation{IBM Quantum, IBM Research Europe -- Zurich, 8803 R\"{u}schlikon, Switzerland}
\author{Francesco Tacchino \orcidlink{0000-0003-2008-5956}}
\email{fta@zurich.ibm.com}
\affiliation{IBM Quantum, IBM Research Europe -- Zurich, 8803 R\"{u}schlikon, Switzerland}

\begin{abstract}
Randomized measurement protocols such as classical shadows represent powerful resources for quantum technologies, with applications ranging from quantum state characterization and process tomography to machine learning and error mitigation.
Recently, the notion of measurement dual frames, in which classical shadows are generalized to dual operators of POVM effects, resurfaced in the literature. This brought attention to additional degrees of freedom in the post-processing stage of randomized measurements that are often neglected by established techniques.
In this work, we leverage dual frames to construct improved observable estimators from informationally complete measurement samples.
We introduce novel classes of parametrized frame superoperators and optimization-free dual frames based on empirical frequencies, which offer advantages over their canonical counterparts while retaining computational efficiency.
Remarkably, this comes at almost no quantum or classical cost, thus rendering dual frame optimization a valuable addition to the randomized measurement toolbox.
\end{abstract}

\maketitle

\section{Introduction}
\label{chap:introduction}
As our abilities to build and operate large quantum information processing systems steadily progress~\cite{kim2023evidence}, the question of finding the most effective strategies to interrogate such devices becomes more and more pressing. 
Indeed, it is well known that to fully reconstruct and store quantum states produced during, say, a quantum simulation, one would need to afford exponentially many physical measurements and a similarly large amount of classical memory.

Luckily, in most practical situations far less than full knowledge about a state is enough to represent meaningful %
information. 
For instance, one may encode a specific problem in the form of a Hermitian operator whose expectation value, evaluated on a carefully optimized quantum state, returns the desired answer~\cite{kandala2017hardwareefficient,cerezo2021variational}. 
This evaluation procedure, also called operator averaging, is much cheaper than full tomography under reasonable sparsity assumptions. 
In practice, however, it can still result in considerable sampling costs~\cite{wecker2015progress,gonthier2022measurements}. 
This motivated the emergence of several protocols to design~\cite{torlai2020precise,garcia-perez_learning_2021,fischer_ancilla-free_2022,glos2022adaptive,jiang2020optimal,huang_predicting_2020,hadfield2022measurements,Hillmich2021,hadfield2021adaptive,huang2021derand,dutt2023practical}, group~\cite{kandala2017hardwareefficient,gokhale_ON3_Measurement_2020,verteletskyi2020measurement,yen2020measuring,hamamura2020efficient,crawford2021efficient,miller2022hardwaretailored,huggins2021efficient,Wu2023overlappedgrouping,yen2023deterministic,Shlosberg2023adaptiveestimation} and schedule~\cite{cotler2020overlapping,garciaperez2020pairwise,bonet2020nearly} optimized sets of quantum measurements over the last few years. 

Among these, approaches based on randomization~\cite{huang_predicting_2020,hadfield2022measurements,elben2022randomized,hu2023scrambled,wan2023matchgate,helsen2022thrifty}, as well as general positive operator-valued measures (POVMs)~\cite{stricker2022experimental,acharya2021}, received substantial attention. 
Upon input of a target quantum state, these strategies return a statistical estimator for it, which is often referred to as a \textit{classical shadow}~\cite{aaronson2018shadow}. 
Although, in general, shadows are not valid quantum states by themselves, they can be efficiently stored and processed, and can be used to reconstruct several incompatible expectation values simultaneously~\cite{huang_predicting_2020}. 
Notably, shadows allow for a separation between the data acquisition phase, which can be carried out without fixing a target observable, and the classical processing and reconstruction stage. 
The power and flexibility of classical shadows led to the development of numerous applications beyond the simple task of operator averaging.
These include the reconstruction of fidelity measures~\cite{Struchalin2021} and of genuine quantum properties of states~\cite{vermersch2023many,joshi2022probing,garcia2021scrambling}, the characterization of quantum processes~\cite{levy2021classical}, classical and quantum machine learning~\cite{huang2022provably,jerbi2023shadows,gyurik2023limitations} and error mitigation techniques~\cite{seif2023shadow,filippov2023scalable, jnane2024error}. 
Often, shadows conveniently serve as a bridge between quantum and classical representations such as tensor networks.

Focusing on the data collection step, several works have considered the optimization of informationally complete (IC) measurement operators to yield estimators with favorable statistical properties~\cite{hadfield2022measurements, huang2021derand, garcia-perez_learning_2021}. 
However, little emphasis has so far been put on studying how the post-processing stage governs the quality of these estimators, particularly of those constructed from overcomplete measurement schemes. 
In fact, it was only in a recent publication that Innocenti et al.~\cite{innocenti2023shadow} raised awareness on this point, highlighting the existence of often neglected degrees of freedom associated with the so-called measurement dual frames~\cite{d2004informationally,kulikov2024minimizing}.
In other words, while classical shadows protocols embody the principle \textit{``measure first, ask questions later''}~\cite{elben2022randomized}, a lot remains to be said about how such questions should be asked.

In this work, we dive deeper into the application of frame theory to IC quantum measurements for digital quantum computing architectures.
We present efficiently computable classes of parametrized dual frames, together with the corresponding optimization routines. 
Indeed, while a number of results are known concerning optimal choices of measurement settings and dual frames for both tomography and observable estimation~\cite{innocenti2023shadow}, these are often impractical or impossible to realize at scale due to inherent technical (e.g., lack of connectivity, device noise) or fundamental (e.g., memory or data processing requirements) limitations. 
By leveraging a product structure, our proposed methods ensure consistent improvements over standard settings while remaining applicable, in principle, up to large sizes of the target qubit registers. 
We support our analytical findings with numerical investigations. 
These suggest that dual frame optimization -- even when subject to certain pragmatic constraints -- can significantly boost the quality of shadow estimators for generic operator averaging tasks. 
In particular, it reduces the performance gap between randomized projective measurements and local dilation POVMs, which are substantially more demanding to implement.

The paper is organized as follows. In Sec.~\ref{chap:theory}, we review the theory of observable estimation with generalized measurements from a frame theory point of view.
In Sec.~\ref{chap:dual_optimization}, we develop methods to optimize dual frame operators to improve the variance of overcomplete POVM estimators.
Finally, in Sec.~\ref{chap:numerics}, we showcase our methods on paradigmatic numerical examples. 

\section{Theory}
\label{chap:theory}
\subsection{Generalized measurements}
\label{sec:generalized_measurements}

The most general class of measurements in quantum mechanics are described by the POVM formalism. 
An $n$-outcome POVM is a set of $n$ positive semi-definite Hermitian operators $\vb{M} = \{M_k\}_{k \in \{1, \dots, n \}}$ that sum to the identity, i.e., $\sum_{k=1}^n M_k = \mathds{1}$. 
Given a $d$-dimensional state $\rho$, the probability of observing outcome $k   $ is given by Born's rule as $p_k = \Tr[\rho M_k]$.
Standard projective measurements (PMs) are a special case of POVMs, where each POVM operator is a projector such that $M_k = \ketbra{\phi_k}{\phi_k}$ for some pure states $\phi_k$.
A POVM is said to be \emph{informationally complete} (IC) if it spans the space of Hermitian operators~\cite{d2004informationally}. 
Then, for any observable $O$, there exist $\w_k \in \mathbb{R}$ such that
\begin{equation}
\label{eqn:observable_POVM_decomp}
O = \sum_{k=1}^{n} \w_k M_k .
\end{equation}
Given such a decomposition of $O$, the expectation value $\expval{O}_\rho$ can be written as
\begin{equation}
\label{eqn:expectation_value_decomp}
\expval{O}_\rho = \Tr[\rho O] = \sum_k \w_k \Tr[\rho M_k] = \mathbb{E}_{k \sim \{p_k\}}[\w_k].
\end{equation}
In other words, $\expval{O}_\rho$ can be expressed as the mean value of the random variable $\w_k$ over the probability distribution $\{p_k\}$.
Given a sample of $S$ measurement outcomes $\{ k^{(1)}, \dots, k^{(S)} \}$, we can thus construct an unbiased Monte-Carlo estimator of $\expval{O}_\rho$ as 
\begin{equation}
\label{eqn:canonical_estimator}
    \hat{o} : \{k^{(1)},\dots, k^{(S)}\} \mapsto \frac{1}{S} \sum_{s=1}^{S} \w_{k^{(s)}}.
\end{equation}

\subsection{PM-simulable POVMs}
\label{sec:PM-simulabel_POVMs}

\begin{figure}
\centering
\includegraphics[width=1\linewidth]{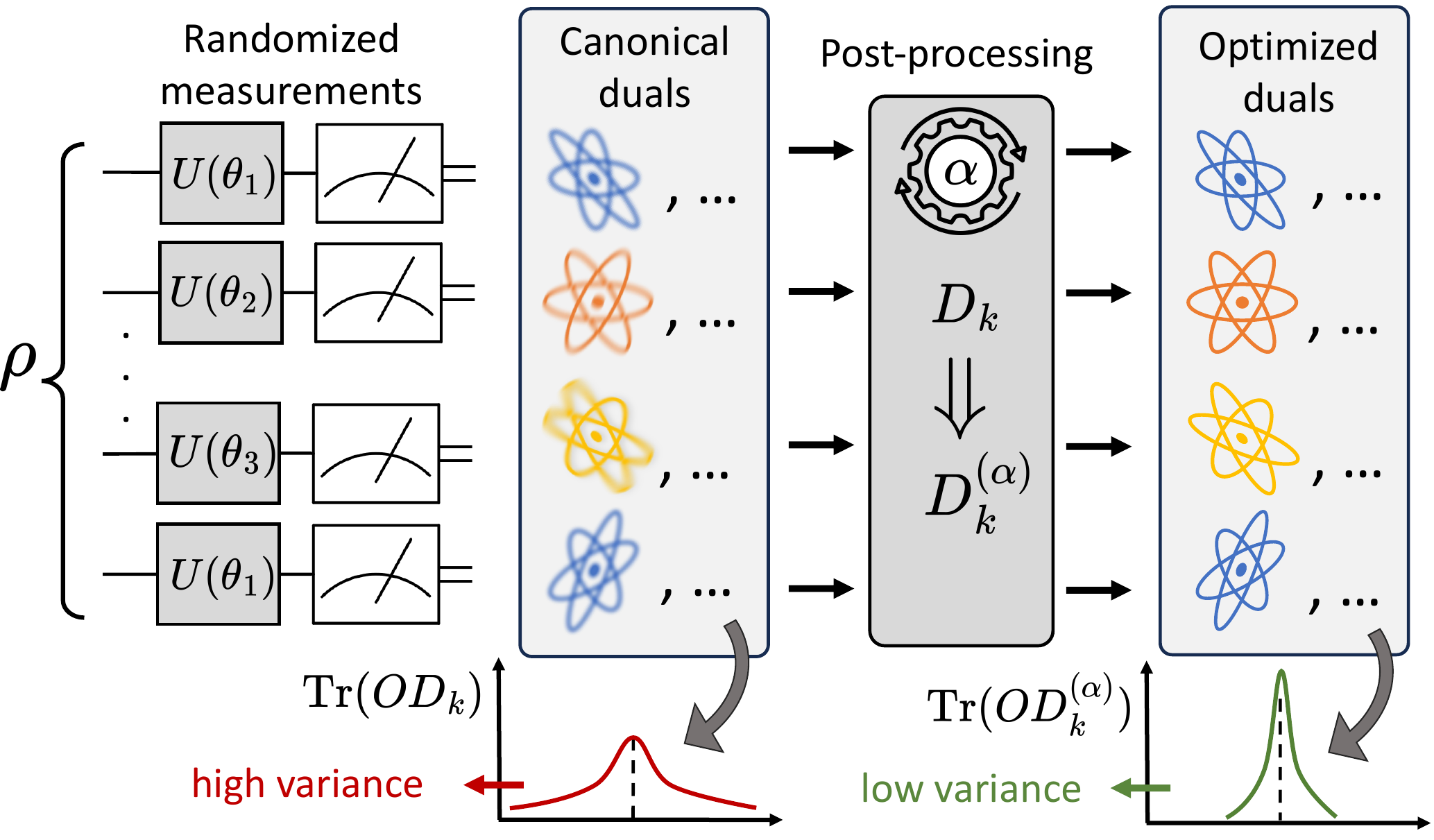} 
\caption[]{
Schematic of dual frame optimization.
 Generalized measurements are performed on the quantum system. 
 Upon obtaining outcome $k$, the corresponding canonical dual operator $D_k$ -- also known as \textit{classical shadow} -- can be efficiently computed and stored on a classical computer. The expectation value of any observable $O$ can be estimated from a sample of dual operators. Leveraging additional degrees of freedom, one can optimize these dual operators through classical post-processing, effectively reducing the estimation variance. 
}
\label{fig:overview}
\end{figure}

Digital quantum computers typically only give access to projective measurements (PMs) in a specified computational basis. 
More general POVMs can be implemented through additional quantum resources, e.g., by coupling to a higher-dimensional space in a Naimark dilation~\cite{gelfand1943imbedding} with ancilla qubits~\cite{chen2007ancilla} or qudits~\cite{fischer_ancilla-free_2022,stricker2022experimental} or through controlled operations with mid-circuit measurements and classical feed-forward~\cite{ivashkov2023highfidelity}.
While these techniques have been demonstrated in proof-of-principle experiments, their full-scale high-fidelity implementation remains a challenge for current quantum devices~\cite{fischer_ancilla-free_2022}.
Of particular interest are thus POVMs that can be implemented without additional quantum resources, i.e., only through projective measurements in available measurement bases.

More complex POVMs can be built from available projective measurements through convex combinations of POVMs:
For two $n$-outcome POVMs $\vb{M}_1$ and $\vb{M}_2$ acting on the same space, their convex combination with elements $M_k = p M_{1,k} + (1-p) M_{2,k}$ for some $p \in [0,1]$ is also a valid POVM. 
This can be achieved in practice by a \emph{randomization of measurements} procedure, which simply consists of the following two steps for each measurement shot.
First, randomly pick $\vb{M}_1$ or $\vb{M}_2$ with probability $p$ or $1-p$, respectively, then perform the measurement associated with the chosen POVM.
We call POVMs that can be achieved by randomizations of projective measurements \emph{PM-simulable}.
On digital quantum computers the easiest basis transformations are single-qubit transformations of the computational basis.
POVMs that consist of single-qubit PM-simulable POVMs are thus the most readily accessible class of generalized measurements and have found widespread application. 
These include classical shadows and most of their derivatives, see Appendix~\ref{app:povm_classes}.

Importantly, PM-simulable informationally-complete POVMs are overcomplete~\cite{dariano_classical_2005}. 
The decomposition of observables from Eq.~\eqref{eqn:observable_POVM_decomp} is thus not unique.
In this work, we leverage these additional degrees of freedom to build better observable estimators, see Fig.~\ref{fig:overview}.

\subsection{Frame theory and dual space}
\label{sec:frames_and_dual_space}
We will now outline a formal approach to obtain the coefficients $\w_k$ in Eq.~\eqref{eqn:observable_POVM_decomp} for a given observable $O$. 
First, we note that the minimal number of linearly independent POVM elements for an IC-POVM is $n = d^2$. 
We call such POVMs \emph{minimally informationally complete}. 
In that case, the coefficients $\w_k$ are unique. 
However, for POVMs with $n>d^2$, such as those that arise from IC PM-simulable POVMs, the decomposition in Eq.~\eqref{eqn:observable_POVM_decomp} is not unique. 
This redundancy is described by frame theory, as outlined in Ref.~\cite{innocenti2023shadow} and detailed in Appendix~\ref{app:frame_theory}.

The set of POVM operators $\vb{M} = \{M_k\}_{k \in \{1, \dots, n \}}$ forms a frame for the space of Hermitian operators if and only if it is IC.
For any frame, there exists at least one dual frame $\vb{D} = \{D_k\}_{k \in \{1, \dots, n \}}$, such that 
\begin{equation} 
\label{eqn:definition_duals}
    O = \sum_{k=1}^n \Tr[OD_k] M_k
\end{equation}
for any Hermitian operator $O$. Therefore, the coefficients $\w_k$ can simply be obtained from the duals $\vb{D}$ as
\begin{equation} 
\label{eqn:coeffs_from_duals}
    \w_k = \Tr[OD_k].
\end{equation}
Notably, dual operators generalize the concept of classical shadows of a quantum state~\cite{huang_predicting_2020}, thus providing a direct connection to the popular randomized measurement toolbox~\cite{elben2022randomized}.

For a minimally IC POVM, only one dual frame exists.
It can be constructed from the POVM elements as 
\begin{equation}
\label{eqn:def_canonical_duals}
\kket{D_k} = \mathcal{F}^{-1} \kket{M_k} \, , \quad k =1,2,\dots,n
\end{equation}
with the \emph{canonical frame superoperator} 
\begin{equation}
\label{eqn:def_frame_superop}
\F = \sum_{k=1}^n \kket{M_k}\bbra{M_k},
\end{equation}
where we have used the widespread vectorized ``double-ket'' notation detailed in Appendix~\ref{app:double-ket_notation}.
Thus, the frame superoperator can be used to transform between the POVM space and the dual space.  
For an overcomplete POVM, the canonical frame superoperator creates one of infinitely many possible dual frames. 
We will further explore this point in section~\ref{sec:parametrization_of_duals}.

\subsection{Observable estimation}
\label{sec:observable_estimation}

The theory of frames and duals enables a systematic approach to estimate observable expectation values from a given set $\{ k^{(1)}, \dots, k^{(S)}\}$ of POVM measurement outcomes:
First, one picks a valid dual frame $\vb{D}$, and construct the dual operators $\{D_{k^{(1)}}, \dots, D_{k^{(S)}}\}$ for the observed outcomes.
Then, one computes the corresponding operator coefficients $\{\w_{k^{(1)}}, \dots, \w_{k^{(S)}}\}$ through Eq.~\eqref{eqn:coeffs_from_duals}.
Finally, Eq.~\eqref{eqn:canonical_estimator} yields an estimate for $\expval{O}_\rho$.
The statistical variance of this estimator is given by the standard error on the mean 
\begin{equation}
\label{eqn:monte_carlo_variance}
    \mathrm{Var}[\hat{o}] = \frac{\mathrm{Var}[\w_{k^{(s)}}]}{S}\, .
\end{equation}
The numerator, also known as the \emph{single-shot variance} (SSV), depends explicitly on the POVM $\vb{M}$, the duals $\vb{D}$ (when they are not unique), the observable $O$ and the state $\rho$ as
\begin{equation} 
\label{eqn:single_shot_variance} \begin{split}
     \mathrm{Var}[\w_k \mid& \, \vb{M}, \vb{D}, O, \rho] = 
    \E{\w_k^2} - \E{\w_k}^2 \\
    & = \sum_k \Tr[\rho M_k] \Tr[O D_k]^2 - \expval{O}_\rho^2.
\end{split} \end{equation}
Throughout this work, the SSV is used as a performance measure for POVM-based estimators. 
Note that the second term $\expval{O}_\rho^2$ depends neither on the POVM nor the dual frame.
However, the first term can be decreased both by adjusting the POVM $\vb{M}$, but also by optimizing the duals $\vb{D}$ (if they are not unique) when the POVM itself remains unchanged, see the schematic in Fig.~\ref{fig:overview}. 
 
The minimal SSV is achieved by performing a PM in the eigenbasis of $O = \sum_k \lambda_k \ketbra{o_k}{o_k}$, in which case $M_k = D_k = \ketbra{o_k}{o_k}$, where $\ket{o_k}$ are the eigenvectors of $O$~\cite{dariano_optimal_2006, hayashi_optimal_2006}.
While this measurement is usually not easily implementable, it serves as a lower bound for all estimations of $\expval{O}_\rho$ with a $\sqrt{S}$-scaling. 
In practice, one typically chooses a specific type of POVM measurement that the quantum hardware can implement, e.g. PM-simulable POVMs or dilation POVMs. 
The POVM operators can then be parametrized and classically optimized to minimize the SSV. 
This can either happen during repeated measurement rounds in an adaptive quantum-classical feedback loop~\cite{garcia-perez_learning_2021, hadfield2021adaptive}, or a priori~\cite{hadfield2022measurements, huang2021derand}. 
Since the SSV depends both on the observable and the generally unknown state, the POVM operators can be optimized either under knowledge of the targeted observable only or by taking into account an approximation of the state obtained from a classical reference calculation.
However, the SSV depends both on the POVM operators as well as the chosen duals whenever these are not unique. 
Crucially, the dependence on the dual frame can be controlled purely during the post-processing phase and thus comes with no additional cost in the quantum resources.
Moreover, an optimization of the dual operators can also be individually tailored to different observables that one might want to estimate from the same set of IC data. 

For a fixed IC-POVM $\vb{M}$ and state $\rho$, the dual frame that minimizes the SSV as a function of $\vb{D}$, irrespective of the observable $O$, is obtained from Eq.~\eqref{eqn:def_canonical_duals} when using a modified frame superoperator given by 
\begin{equation}
\label{eqn:optimal_dual_frame} 
\F_\text{opt} = \sum_{k=1}^n \frac{1}{\Tr[\rho M_k]}\kket{M_k}\bbra{M_k}
\end{equation}
and re-scaling each dual operator by $1/\Tr[\rho M_k]$~\cite{innocenti2023shadow}.
This, however, requires knowledge of the state $\rho$.
If no prior information on $\rho$ is available, the best choice of duals can be considered the one that minimizes the SSV as a uniform Haar average over all states. 
In that case, the optimal duals are obtained from a modified frame superoperator given by $\F_\text{avg} = \sum_{k=1}^n \kket{M_k}\bbra{M_k} / \Tr[M_k]$. 
It is important to point out that this choice is the one generally employed in classical shadows protocols~\cite{elben2022randomized}.
More explicitly, a standard classical shadow $\hat \rho_s$, namely a single-shot estimate of the state obtained by constructing an inverse measurement channel, is equivalent to the ``average-optimal'' dual $D_k$ obtained from inversion of the frame superoperator $\F_\text{avg}$ introduced above.  

\section{Dual frame optimization}
\label{chap:dual_optimization}
\subsection{Local POVMs and duals}
A POVM-based estimation of observables is only feasible if the POVM operators themselves as well as the dual operators can be efficiently handled classically. 
Therefore, most POVM-based estimation schemes employ local POVMs, where every $M_k$ acts non-trivially only on few qubits. 
This is crucial in order to keep a local structure in the frame superoperator from Eq.~\eqref{eqn:def_frame_superop}, such that it can be constructed and inverted efficiently. 
Moreover, also the duals themselves need to be efficiently processable in order to compute the observable coefficients $\w_k$ via Eq.~\eqref{eqn:observable_POVM_decomp}. 
However, this is not guaranteed by the optimal dual frame $\F_\text{opt}$, even when the POVM operators have a product structure. 
This implies that the optimality results presented in the previous chapter cannot be applied in general.

In the following we consider $N$-qubit systems where the POVM effects are tensor products of single-qubit $n$-outcome POVM effects. That is, each global effect can be written as
\begin{equation}
\label{eqn:povm_product_form}
    M_{\vb{k}} =
    M_{k_1,k_2,\dots,k_N} =
    M^{(1)}_{k_1} \otimes M^{(2)}_{k_2} \otimes \cdots \otimes M^{(N)}_{k_N}  
\end{equation}
where $\vb{M}^{(i)} = \{ M^{(i)}_{k_i} \}_{k_i=1}^{n}$
is a $n$-outcome single-qubit POVM acting on qubit $i$. 
To fully leverage the ability of optimizing duals for the SSV, we require a parametrization of suitable dual frames, such that they remain efficiently processable.  
We provide this through a parametrized frame superoperator in the following. 

\subsection{Parametrization of duals}
\label{sec:parametrization_of_duals}

\subsubsection{Weighted frame superoperator}
\label{sec:weighted_frame_superop}
If $\vb{M}$ is overcomplete, the set of all valid duals can be explicitly parametrized through a singular value decomposition~\cite{krahmer_sparsity_2013}, see Appendix~\ref{app:svd_param_duals}.
In principle, this parametrization could be used to optimize the dual frame for a minimal SSV. 
However, it is not straightforward to impose a product structure on the dual operators in this way.
For a more practical, albeit non-exhaustive parametrization of the dual frames, we thus define a \emph{weighted frame superoperator} 
\begin{equation} 
\label{eqn:alpha_weighted_framesuperop}
    \mathcal{F}_\alpha = \sum_{\vb{k}} \alpha_{\vb{k}} \kket{M_{\vb{k}}} \bbra{M_{\vb{k}}} 
\end{equation}
which resembles the canonical frame superoperator $\F$, but with the contribution of each effect $M_{\vb{k}}$ rescaled by a factor $\alpha_{\vb{k}}\in \mathbb{R}$.
If $\mathcal{F}_\alpha$ is invertible, the effects given by
\begin{equation} 
\label{eqn:alpha_weighted_duals}
    \kket{D_{\vb{k}}} = \alpha_{\vb{k}} \mathcal{F}_\alpha^{-1} \kket{M_{\vb{k}}}
\end{equation}
form a valid dual frame which is invariant under uniform scaling of the coefficients.
Notice that if all the coefficients $\{ \alpha_{\vb{k}} \}$ are positive, then $\mathcal{F}_\alpha$ will be positive definite and hence invertible.  
We can hence think of the parameters $\{ \alpha_{\vb{k}} \}$ as a probability distribution when restricting them to positive values.

Assuming the POVM has a product structure as in Eq.~\eqref{eqn:povm_product_form}, it is the degree of correlations in the multivariate probability distribution $\alpha_{\vb{k}}$ that determines what kind of product structure the resulting duals from Eq.~\eqref{eqn:alpha_weighted_duals} will have.
In the simplest case, when $\alpha_{\vb{k}}$ fully factorizes, i.e., $\alpha_{k_1, k_2, \dots, k_N} = \alpha^{(1)}_{k_1} \alpha^{(2)}_{k_2} \cdots \alpha^{(N)}_{k_N}$, the dual frame will be of product form as in Eq.~\eqref{eqn:povm_product_form}. 
More generally, if $\alpha_{\vb{k}}$ is a product of distributions that each act on at most $m$ qubits, then the duals will be tensor products of terms that act on $m$ qubits. 
In this case, the traces to compute $\omega_k$ in Eq.~\eqref{eqn:coeffs_from_duals} factorize into blocks that involve constructing matrices of at most size $2^m \times 2^m$. 
This way, the complexity of the dual operators in the post-processing can be tuned by imposing restricted correlations in the distribution $\alpha_{\vb{k}}$.

We thus propose the following general procedure to improve statistical estimators based on overcomplete POVMs such as classical shadow methods. 
First, a collection of shots $\{ k^{(1)}, \dots, k^{(S)} \}$ is measured from a fixed POVM.  
For a given parametrization of duals through equation~\eqref{eqn:alpha_weighted_duals}, the single-shot variance in Eq.~\eqref{eqn:single_shot_variance} is estimated  with the (corrected) sample variance of the values $\{\omega_{k^{(1)}}, \dots, \omega_{k^{(S)}} \}$.
An optimizer will then minimize the SSV as a function of the parameters entering the weighted frame superoperator, yielding an estimator of $\langle O \rangle_\rho$ with the smallest possible variance. 
This can be repeated independently for each observable of interest starting from the same collection of samples, harnessing the true power of IC measurements.
As this dual optimization does not require changing the quantum circuits to be executed nor increasing the sample size, we consider it to be a ``free lunch'' improvement over standard classical shadows techniques. 

In practice, the optimization landscape of the dual parameters $\alpha_{\vb{k}}$ could be difficult to navigate, due to the complicated dependency of the duals on $\alpha_{\vb{k}}$ in Eq.~\eqref{eqn:alpha_weighted_duals}.
Also, a simultaneous optimization of the duals and the POVM operators themselves can be cumbersome, as it requires a quantum-classical feedback loop. 
As an alternative to optimizing a parametrization of the dual operators, we thus propose the following procedure to obtain suitable dual frames for a fixed and overcomplete IC POVM, which we refer to as \emph{empirical frequencies dual frames}. 

\subsubsection{Empirical frequencies dual frames}
\label{sec:empirical_frequency_dual_frames}
When no knowledge about the state is available, the average-optimal dual frame should be used, as discussed in Sec.~\ref{sec:observable_estimation}.
As the POVM measurement is repeated and the number of shots $S$ increases, we gain some knowledge about the state, which we can leverage to approximate the optimal dual from Eq.~\eqref{eqn:optimal_dual_frame}.
More precisely, the measured frequencies $f_{\vb{k}} = \#{\vb{k}} / S$ (where $\# {\vb{k}}$ is the number of times the outcome ${\vb{k}}$ was obtained) follow a multinomial distribution and converge to the true measurement probabilities $p_{\vb{k}} = \Tr[\rho M_{\vb{k}}]$ as $\sqrt{p_{\vb{k}}(1-p_{\vb{k}})/S}$.
One could thus replace the outcome probabilities $p_{\vb{k}} = \Tr[\rho M_{\vb{k}}]$ in the optimal dual frame with the empirical frequencies $f_{\vb{k}} = \#{\vb{k}} / S$. That is, we use the \emph{global empirical dual frame}
\begin{equation} \label{eq:empirical_dual_frame} \begin{split}
    \kket{D_{\vb{k}}} &= \frac{1}{f_{\vb{k}}} \mathcal{F}^{-1}\kket{M_{\vb{k}}} \\
    \textrm{where } \mathcal{F} &= \sum_{\vb{k}} \frac{1}{f_{\vb{k}}} \kket{M_{\vb{k}}}\bbra{M_{\vb{k}}}
\end{split} \end{equation}
which is a weighted frame superoperator with $\alpha_k = 1/f_k = S/\# k$.
However, an obvious issue arises if an outcome is not obtained ($f_k =0$).
We address this by adding a regularization to the empirical frequencies with a bias term.
This biases the outcome probabilities with respect to the fully mixed state $\frac{1}{d}\mathds{1}$, borrowing an idea from Ref.~\cite{hayashi_optimal_2006}. 
The resulting \emph{biased empirical frequencies} are given by 
\begin{equation}
\label{eqn:biased_empirical_frequencies}
    \Tilde{f}_{\vb{k}}(\{k^{(1)},\dots,k^{(S)}\},S_{\textrm{bias}}) = \frac{\# {\vb{k}} + \Tr[\frac{1}{d}\mathds{1} M_{\vb{k}}] S_{\textrm{bias}}}{S + S_{\textrm{bias}}} \, .
\end{equation}
If we assume that all effects are non-null, then $\alpha_{\vb{k}} = 1/\Tilde{f}_{\vb{k}} >0 $, which ensures the frame superoperator is invertible.
Note that, for $S=0$, we recover the average-optimal dual frame, while for $S \rightarrow \infty$ the empirical dual frame converges to the optimal dual frame from Eq.~\eqref{eqn:optimal_dual_frame}.

The global empirical dual frame still suffers from two issues: 
Firstly, for sizable qubit numbers, the number of different POVM outcomes $n$ eventually becomes much larger than the available shot budget $S$. 
In this regime, it is difficult to improve over the average-optimal dual with the above global empirical dual frame.
Secondly, the dual matrices can become exponentially large when the correlations in $\Tilde{f}_{\vb{k}}$ are not restricted, as discussed in Sec.~\ref{sec:weighted_frame_superop}. 
Both of these issues are overcome when relaxing the task from learning the global distribution $p_{\vb{k}}$ to recovering only the most relevant few-qubit correlations of this multivariate distribution. 
In the simplest case, the (potentially biased) outcome probabilities $f_{\vb{k}}$ can be approximated with the product of marginal probabilities 
\begin{equation}
\mathfrak{f}_{\vb{k}} = \prod_{i=1}^N \mathfrak{f}^{(i)}_{k_i} \, , \;\;\; \textrm{with} \;\;\; \mathfrak{f}^{(i)}_{k_i} = \sum_{\{k_j\}_{j\neq i}} f_{k_1, \dots, k_N} \, .
\end{equation}
While this does not model correlations between POVM outcomes of different qubits, it still presents an advantage over the average-optimal dual frame, while ensuring the dual frame is of product form, see Sec.~\ref{chap:numerics}.

The correlations captured by the empirical frequencies can be systematically tuned up by partitioning $\vb{k}$ into marginals of larger sizes. 
Let $\Lambda = \{ \lambda_1, \dots, \lambda_l \}$ be a partitioning of the qubit indices $\{1,\dots, N\}$ into subsets $\lambda_i$ that each contain up to $m$ terms. 
We can then approximate the global distribution $f_{\vb{k}}$ (or $\tilde f_{\vb{k}}$)  as a product of \emph{$m$-body marginals} $\mathfrak{f}_{\lambda_i}$
\begin{equation}
\label{eqn:m-body-marginals}
\mathfrak{f}_{\vb{k}}^{\Lambda} =  \prod_{i=1}^l \mathfrak{f}_{\lambda_i} \, , \;\;\; \textrm{with }  \;\;\; \mathfrak{f}_{\lambda_i} = \sum_{\{k_j\}_{j\notin \lambda_i}} f_{k_1, \dots, k_N} \, .
\end{equation}
This leads to dual frame operators that are tensor products of $m-$local terms. 
The question arises how to optimally choose the partitioning $\Lambda$.
Ideally, pairs of qubits whose POVM outcomes are highly correlated should preferably be grouped into the same set. 
We quantify this through the empirical mutual information $I(i, j)$ shared by two qubits, given as 
\begin{equation}
\label{eqn:mutual_information}
I\left( i, j \right) = \sum_{k_i, k_j} \mathfrak{f}_{\{i, j\}} \log \left(\frac{ \mathfrak{f}_{\{i, j\}}}{   \mathfrak{f}^{(i)}_{k_i} \mathfrak{f}^{(j)}_{k_j}} \right).
\end{equation}
This quantifies the price one has to pay when approximating the joint distribution $\mathfrak{f}_{\{i, j\}}$ through the product of marginal distributions $ \mathfrak{f}^{(i)}_{k_i}  \mathfrak{f}^{(j)}_{k_j}$, given by their Kullback-Leibler divergence.

In a practical setting, the maximally-allowed degree $m_\text{max}$ should be chosen such that the classical cost in computing the traces of the resulting $2^m \times 2^m$ dual matrices is deemed tolerable, and sufficient statistics are gathered to capture the $m-$body marginals, which becomes exponentially more difficult as $m$ increases.
Once $m_\text{max}$ is chosen, one can define a cost function $\mathcal{C}$ for $\Lambda$
that is given by the sum of the pairwise mutual information over all set, i.e., 
\begin{equation}
\label{eqn:partitioning_score}
\mathcal{C}(\Lambda) = \sum_{\lambda_i \in \Lambda} \sum_{\substack{j, k \in \lambda_i \\ j \neq k}} I(j, k).
\end{equation}
While the optimal partitioning can be straightforwardly computed from this cost function for small qubit numbers, this becomes infeasible for larger $N$ due to the super-polynomial scaling of the number of different partitionings. 
In such a setting, one can construct a well-performing partitioning by first computing $I(i, j)$ for all pairs of qubits, and then putting pairs of highest values into the same subset with a greedy allocation strategy.

\section{Numerical benchmarks}
\label{chap:numerics}
\begin{figure*}
\centering
\includegraphics[width=0.99\linewidth]{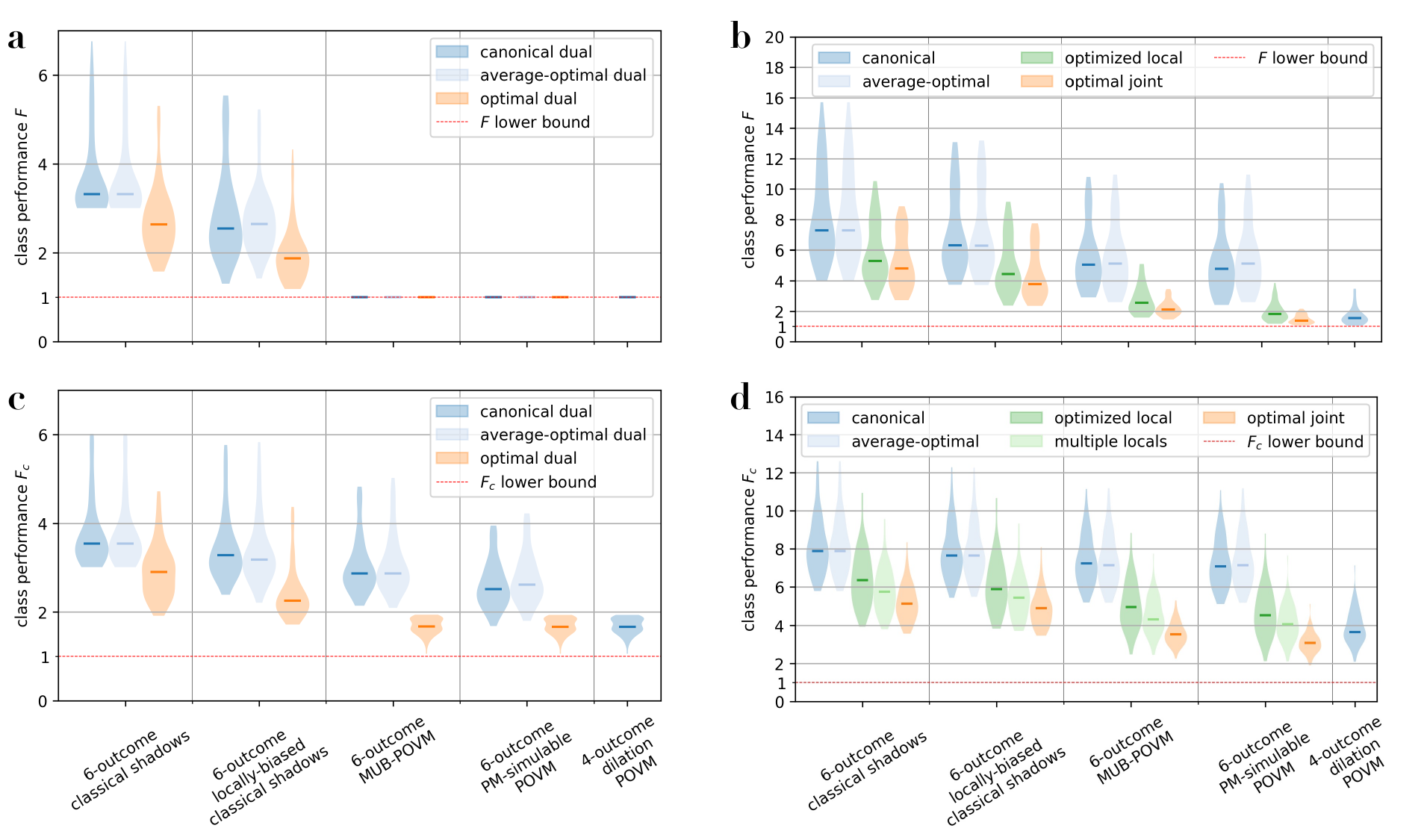} 
\caption[]{
Performance of different classes of POVMs and dual frames for estimating random observables, shown as violin plots. 
Color indicates the employed class of duals.
Each distribution is built from 200 repetitions of sampling a Haar-random pure state and an observable (or set of observables) and subsequently minimizing the class performance $F$ (or cumulative performance $F_\text{C}$) for each combination of POVM and dual frame.
Horizontal markers show the median of each distribution.
For blue distributions a fixed dual frame is used. 
For orange distributions, the optimal dual frame is used.
The red dashed line represents the optimal lower bound which is saturated for the projective measurement in the eigenbasis of the observable.
\textbf{a)} Single-qubit system with one observable.
\textbf{b)} Two-qubit system with one observable. 
\textbf{c)} Single-qubit system with five observables. 
\textbf{d)} Two-qubit system with five observables.
For two-qubit systems, the class performance is computed for the cumulative variance of all observables and the optimized dual frames are limited to product form for the green distributions. 
For the light-green distributions, the duals are re-optimized for every observable. 
}
\label{fig:povm_class_performances}
\end{figure*}

We now showcase the methods for dual frame optimization developed in Sec.~\ref{chap:dual_optimization} on numerical examples. 
In the following, in Sec.~\ref{sec:performance_limits_numerics}, we first benchmark the performance of different classes of POVM operators and dual frames by optimizing their single-shot variance for generic random states and observables.
Then, in Sec.~\ref{sec:numerical_results_empirical_frequencies}, we demonstrate how the explicit optimization of the operators can be circumvented by using only the empirical frequencies of the outcomes to obtain a well-performing dual frame.

\subsection{Performance limit of different POVM classes}
\label{sec:performance_limits_numerics}

Here, we investigate which class of single-qubit POVMs and dual frames yields the best possible estimators. 
In this idealized setting, we assume full knowledge of the underlying state $\rho$. 
Our procedure is the following: 
We first sample a Haar-random pure state $\rho$. 
We construct random observables $O$ by sampling eigenvalues $\lambda_1, \dots, \lambda_d$ uniformly at random from 
$\left[-5, 5\right]$ and applying a Haar-random unitary $U$ that yields $O = U \mathrm{diag}(\lambda_1, \dots, \lambda_d) U^\dagger$.
As discussed in Sec.~\ref{sec:observable_estimation}, the optimal measurement would be the projective measurement in the eigenbasis with a SSV of $\expval{O^2}_\rho - \expval{O}^2_\rho$.
For a given observable $O$, we therefore evaluate a class of POVMs $\mathcal{M}$ and duals $\mathcal{D}$ by their \emph{class performance}
\begin{equation}
\label{eqn:performance_limit}
    F(\mathcal{M}, \mathcal{D}) = \ \min_{\vb{M} \in \mathcal{M}, \vb{D} \in\mathcal{D} } \left\{\frac{ \mathrm{Var}[\w_k \mid \vb{M}, \vb{D}, O, \rho] }{  \expval{O^2}_\rho - \expval{O}^2_\rho}\right\},
\end{equation}
with $\mathrm{Var}[\w_k]$ given as in Eq.~\eqref{eqn:single_shot_variance}. 
Similarly, we quantify the ability to estimate several observables $\{O_i\}_{i \in \{1,\dots, N_\text{obs}\}}$ from the same IC POVM data through the \emph{cumulative class performance}
\begin{equation}
\label{eqn:cumulative_performance_limit}
    F_\text{C}(\mathcal{M}, \mathcal{D}) = \min_{\substack{\vb{M} \in \mathcal{M} \\ \{\vb{D}_i\}_{i \in \mathcal{D} }}} \left\{
    \frac{\sum\limits_{i=1}^{N_\text{obs}}  \mathrm{Var}[\w_k \mid \vb{M}, \vb{D}_i, O_i, \rho] }{ \sum\limits_{i=1}^{N_\text{obs}} \expval{O_i^2}_\rho - \expval{O_i}^2_\rho}
    \right\}.
\end{equation}
Note that the duals $\vb{D}$ in Eqs.~\eqref{eqn:cumulative_performance_limit} and~\eqref{eqn:performance_limit} are implicitly defined through the POVM operators $\vb{M}$ but carry free parameters as in Eq.~\eqref{eqn:alpha_weighted_duals}.
We use a BFGS optimizer to compute $F$ and $F_\text{C}$ for five classes of single-qubit POVMs detailed in Appendix~\ref{app:povm_classes}, namely, classical shadows, locally-biased classical shadows, mutually-unbiased bases (MUB) POVMs, and general PM-simulable POVMs (all 6 outcomes each), as well as 4-outcome dilation POVMs.
The distributions of the achieved performance limit for 200 random states $\rho$ state and observables $O$ (or set of $N_\text{obs}=5$ observables $\{O_i\}$ for $F_C$ ) are shown in Fig.~\ref{fig:povm_class_performances} for single-qubit and two-qubit systems. 

In all cases, the class performance improves significantly when moving beyond the canonical dual frame. 
Therefore, the additional degrees of freedom leveraged by our dual optimization improve POVM estimators beyond what can be achieved by optimizing the POVM operators alone. 
Trivially, on a single-qubit observable, the MUB, PM-simulable and dilation POVMs all reach the optimal performance $F=1$, as the eigenbasis projectors are included in this class of POVMs, see Fig.~\ref{fig:povm_class_performances}\textbf{a}. 
However, when estimating several observables from the same POVM data, the cumulative performance is again improved by adapting the dual frame for each observable, see Fig.~\ref{fig:povm_class_performances}\textbf{c}. 
For two-qubit observables, no measurement setting will consistently reach the optimal performance as we restrict ourselves to single-qubit POVM operators, see Fig.~\ref{fig:povm_class_performances}\textbf{b}. 
The optimized local duals perform slightly worse than the optimized global duals but still considerably better than canonical duals. 
When estimating several two-qubit observables, optimizing the dual operators gives a more significant performance improvement than adding more complexity to the measured POVM operators by going from classical shadows to more general PM-simulable POVMs, see Fig.~\ref{fig:povm_class_performances}\textbf{d}.
The optimal local dual frame depends on the observable, hence it should be re-optimized it for every observable, offering an additional slight improvement.

A common trend in the results of Fig.~\ref{fig:povm_class_performances} is the following: As more degrees of freedom are optimized in a PM-simulable POVM, the performance gains become increasingly smaller and reach a plateau when using the canonical dual frame (see, e.g., blue violins in Fig.~\ref{fig:povm_class_performances}\textbf{d}). 
However, these gains become increasingly larger when using optimized dual frames. 
In other words, it becomes less and less worth it to add further degrees of freedom to the POVM operators when using the canonical dual frame, which is the opposite to what is observed when using optimized dual frames.
This is especially true when estimating several observables from the same POVM data.
In all cases, PM-simulable POVMs with optimized duals (even local ones) come close to or surpass the performance of optimized dilation POVMs. 
Interestingly, the average-optimal dual frame (see Sec.~\ref{sec:observable_estimation}) does not offer reliable performance improvements, indicating that this result might not be practical in a realistic setting.

\subsection{Empirical frequencies dual frames}
\label{sec:numerical_results_empirical_frequencies}

\begin{figure}
\centering
\includegraphics[width=0.99\linewidth]{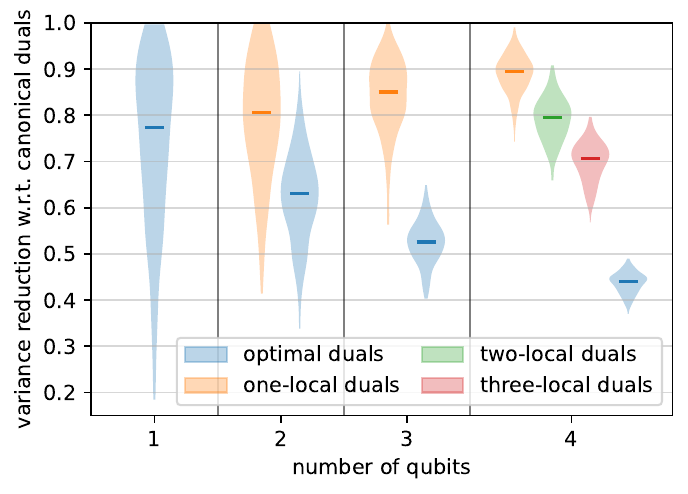} 
\caption[]{
Variance reduction compared to classical shadows with canonical duals for different types of empirical frequencies dual frames. 
Violin plots show the distribution over 200 random pairs of states and observables.
}
\label{fig:marginals_scaling_with_qubit_number}
\end{figure}

Next, we showcase how to bypass the explicit optimization of the dual frame with the use of empirical frequencies dual frames as introduced in Sec.~\ref{sec:empirical_frequency_dual_frames}.
We first investigate the performance of the $m$-body marginal distributions in the infinite shot limit, i.e., we construct the marginal probabilities in Eq.~\eqref{eqn:m-body-marginals} from the exact outcome distributions $p_{\vb{k}}$. 
In Fig.~\ref{fig:marginals_scaling_with_qubit_number}, we show how the improvement over canonical duals scales with an increase in the system size. 
Here, we plot the ratio of the variance of classical shadows estimators when using optimized duals compared to canonical duals.
Distributions in violin plots are obtained from 200 random samples of states and (single) observables. 
Remarkably, as the system size increases from one to four qubits, the improvement of the optimal global duals becomes more and more pronounced. 
At four qubits, the variance of this optimal estimator is less than half of the variance of the canonical dual estimator. 
This can be understood as a consequence of increasing the classical resources that go into the construction of the dual operators. 
The duals derived from the single-qubit marginal distributions (one-local duals) become less performant as the system size increases. 
This comes as no surprise, as the product of the marginal distributions will capture the true correlated distribution less and less successfully with increasing qubit number.
The performance of the marginal duals can be systematically improved by including higher-order correlations, as shown by the two-local and three-local duals. 
These are constructed by choosing the optimal partitioning of the four qubits into subsets of sizes $(2,2)$ and $(3,1)$ according to Eq.~\eqref{eqn:partitioning_score}. 
Overall, these numerical results show that marginal frequencies dual frames can offer a straightforward improvement over canonical dual frames. 
Their performance can be systematically boosted by introducing higher-order correlations.

\begin{figure}
\centering
\includegraphics[width=1\linewidth]{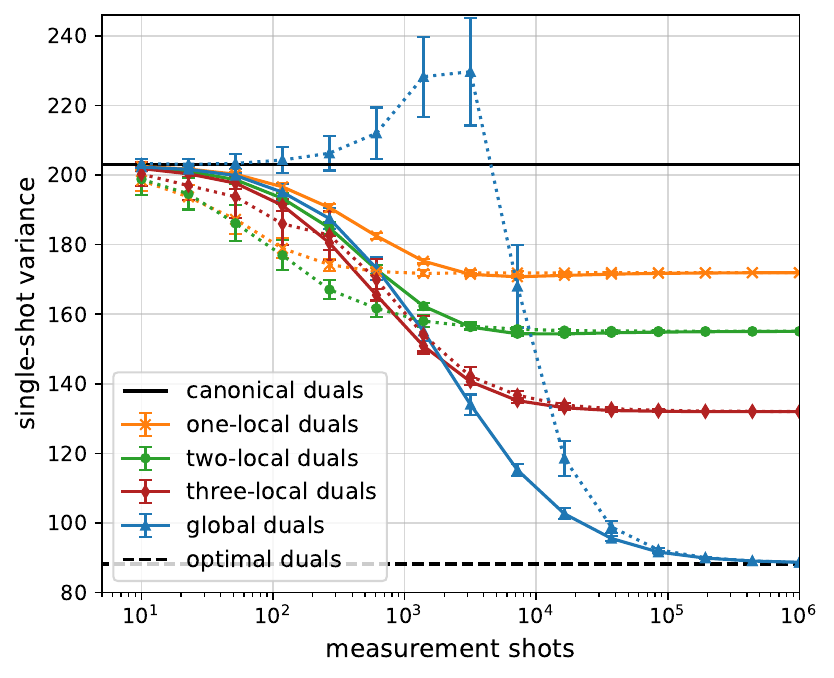} 
\caption[]{
Convergence of the single-shot variance with increasing shot number for estimators based on empirical frequency dual frames on random four-qubit states and observables.
The POVM operators are single-qubit classical shadows. 
For the solid line data, a bias of $S_\text{bias} = 1296$ is used, while dashed lines are obtained with $S_\text{bias} = 128$.
Error bars are the standard deviation over 15 repetitions of sampling the indicated number of shots from the underlying POVM distribution. 
}
\label{fig:marginals_shots_scaling}
\end{figure}

Finally, we investigate how well empirical frequencies perform in the practically relevant setting of finite samples.
In Fig.~\ref{fig:marginals_shots_scaling}, we show how the SSV improves with increasing sample size $S$ for different types of marginal distributions when estimating a single four-qubit observable.
Note that we use the biased empirical frequencies introduced in Eq.~\eqref{eqn:biased_empirical_frequencies} to choose the dual frame, 
but plot the true underlying SSV according to Eq.~\eqref{eqn:single_shot_variance} instead of estimating the variance from the finite sample.
To illustrate the role of $S_\text{bias}$, we show the convergence with one lower value of $S_\text{bias}=128$ and one larger value of $S_\text{bias}=1296$ (the number of different POVM outcomes). 
In the regime where $S\leq S_\text{bias}$, the duals remain close to the canonical duals by design. 
As $S$ increases, the empirical frequencies eventually converge to their true underlying values.
The bias controls the rate and stability of this statistical convergence.
The smaller bias is sufficient to give a smooth convergence for the more restricted marginal distributions (dotted lines). 
In fact, the empirical frequencies with the one-local, two-local, and three-local marginals already offer a concrete improvement over the canonical dual variance with only a few hundred measurement samples, which is well below the total number of POVM outcomes. 
However, when approximating the global outcome distribution (blue curves), choosing too small of a bias will render the empirical frequencies unstable (dashed blue line). 
On the other hand, for the one-local and two-local duals, the larger bias comes at the price of a significantly slower convergence, illustrating a tradeoff between stability and speed. 
In practice, $S_\text{bias}$ can nevertheless always be chosen large enough such that the empirical frequencies dual frame gives a performance that is at least as good as the canonical duals.
Our results indicate that a reasonable choice for $S_\text{bias}$ is on the order of the degrees of freedom in the marginalized probability distributions. 

\section{Discussion}
\label{chap:discussion}
In this paper, we investigated the connection between frame theory and randomized overcomplete quantum measurements~\cite{innocenti2023shadow}. 
In particular, we developed and tested scalable dual frame optimization strategies that allow us to significantly sharpen the performances of ``measure-first'' schemes while acting entirely at the data post-processing stage. 

Inspired by known analytical results which, albeit optimal, are hardly realizable in practice, we identified a minimal set of constraints that guarantee an efficient and effective computational pipeline. 
More specifically, we proposed a marginalized version of dual frames offering advantages even if restricted to a structure of limited correlations.
This class of duals can, in principle, be parameterized and iteratively refined in combination with, e.g., adaptive POVMs~\cite{garcia-perez_learning_2021}. 
Furthermore, we described an optimization-free dual frame, based on empirical outcome frequencies, that converges (in the statistical sense) to the best $m$-local one. 
This approach does not require any prior knowledge or assumptions on the state being measured, and can be tuned up systematically to capture the most relevant correlations in the measurement outcomes of individual qubits.
By removing the need to explicitly optimize the dual frame, this solution significantly simplifies the navigation of the measurements space when searching for the most suitable POVM/dual combination. Our techniques are especially relevant for use-cases that require estimating several observables for the same state, as the dual operators can be optimized for every observable independently. Beyond reconstructing expectation values, all our proposed methods are applicable to an extensive set of tasks, including reduced state tomography, machine learning. and error mitigation.

To support our analysis, we performed numerical simulations in both the infinite statistics limit and in the finite sample regime. Remarkably, our results suggest that, with a judicious selection of the dual frame, overcomplete PM-simulable POVMs can come close to the best results obtained with dilation ones in the context of operator estimation. Due to the inherent simplicity of their implementation, PM-simulable POVMs might then be preferable in practical applications.
Our simulations -- albeit collected at modest scales and for generic, Haar-random states and observables -- also indicate that the advantage unlocked by using global duals increases with the system size, while the relative improvement brought by simple single-qubit marginal dual frames decreases. 
We note that, in most practical settings, observables are seldom dense and are rather linear combinations of Pauli strings with finite weight, e.g.\ Hamiltonians encountered in condensed matter, lattice problems and quantum chemistry.
For such local Pauli strings, one can construct a dual that acts globally on the qubits that make up the non-trivial support of the observables. Furthermore, physically relevant states most often exhibit a distinctive and restricted correlation structure, which can be exploited in the construction of our proposed mutual-information-based dual frames. 
We leave the in-depth study of these aspects to future work.

As a next step, one could aim at better characterizing the gap between ideal and marginal dual frames in a systematic way, possibly using tools from probability theory~\cite{watanabe_information_1960,fano1961}. 
In parallel, it would also be interesting to focus on the development of alternative classes of efficiently computable dual frames which could help reduce such performance separation. 
Examples might draw inspiration from Clifford~\cite{helsen2022thrifty,hu2023scrambled} and matchgate~\cite{zhao2021fermionic,wan2023matchgate} shadows, or make use of classical techniques such as tensor networks~\cite{filippov2023scalable} and neural network quantum states~\cite{torlai2018neural}. 
Combining estimates obtained with different dual frames, for example through a median of means~\cite{huang_predicting_2020}, could also improve the overall quality of the estimators.

To conclude, and as a key takeaway, our work confirms that dual frames deserve a greater attention whenever overcomplete IC techniques are employed to reconstruct properties of quantum states. 
In fact, the freedom they offer only concerns the classical processing of outcomes 
and can be straightforwardly leveraged to improve any shadow-based protocol without overheads in sampling or quantum circuit complexity. 
We therefore expect that the careful selection of duals will soon become a standard component of randomized and IC measurement toolboxes, significantly enhancing their performances.

\textit{Note added}. Recently, we became
aware of related work by J.~Malmi \textit{et.~al}~\cite{malmi2024enhanced}, in which the authors report estimation improvements using dual optimization in the context of Hamiltonian simulation and quantum chemistry. A.~Caprotti \textit{et.~al}~\cite{caprotti2024optimising} also report complementary results concerning optimised shadow inversion maps.

\section*{Acknowledgments}
We thank David Sutter for fruitful discussions. 
This research has received funding from the European Union’s Horizon 2020 research and innovation program under the Marie Sk\l{}odowska-Curie grant agreement No.~955479 (MOQS – Molecular Quantum Simulations). 
This research was supported by the NCCR MARVEL, funded by the Swiss National Science Foundation.
IBM, the IBM logo, and ibm.com are trademarks of International Business Machines Corp., registered in many jurisdictions worldwide. Other product and service names might be trademarks of IBM or other companies. 
The current list of IBM trademarks is available at \url{https://www.ibm.com/legal/copytrade}.

\clearpage
\appendix

\section{Classes of single-qubit IC POVMs}
\label{app:povm_classes}
In this appendix, we detail the classes of single-qubit IC POVMs used for numerical benchmarks in Sec.~\ref{sec:performance_limits_numerics}.
In Sec.~\ref{sec:PM-simulabel_POVMs}, we introduced the concept of convex combinations of POVMs that can be implemented through classical randomization of the involved POVMs.
Here, we build on this concept to construct POVMs of increasing complexity. 
We use the notation 
\begin{equation}
\label{eqn:multiset_sum}
 \biguplus_i q_i \vb{M}_i = \left\{ q_i M_{i,k} \right\}_{i,k} 
\end{equation}
to denote a convex combination of POVMs $\vb{M}_i$ acting on the same Hilbert space, such that the POVM $\vb{M}_i$ is implemented with probability $q_i$.

\subsection{Classical shadows}
Let $\vb{P}_Z = \{\ketbra{0}{0}, \ketbra{1}{1} \}$ be the computational basis measurement.
Under transformation of single-qubit Clifford gates $\{H, S\}$ (with the Hadamard gate $H$ and the phase gate $S$), we get access to the projective measurements in the $X$ and $Y$ basis, i.e., $\vb{P}_X = \{\ketbra{+}{+}, \ketbra{-}{-} \}$ and $\vb{P}_Y = \{\ketbra{+i}{+i}, \ketbra{-i}{-i} \}$. In their simplest formulation, \emph{classical shadows} protocols~\cite{huang_predicting_2020} implement the POVM 
\begin{equation}
    \frac{1}{3} \vb{P}_X \uplus \frac{1}{3} \vb{P}_Y \uplus \frac{1}{3} \vb{P}_Z
\end{equation}
by measuring a given qubit in one of these Pauli bases with equal probability.

\subsection{Locally-biased classical shadow}
Classical shadows can be generalized by treating the probabilities of measuring a qubit with $\vb{P}_X$, $\vb{P}_Y$, or $\vb{P}_Z$ as additional degrees of freedom~\cite{hadfield2022measurements}.
We refer to the ensuing single-qubit POVM
\begin{equation}
    q_X \vb{P}_X \uplus q_Y \vb{P}_Y \uplus q_Z \vb{P}_Z
\end{equation}
as \emph{locally-biased classical shadows} (LBCS).
This POVM carries two free parameters $q_X$ and $q_Y$, with $q_Z = 1 - q_X - q_Y$. 

\subsection{MUB-POVM}

Consider a set of different orthonormal bases for a $d$-dimensional Hilbert space.
These bases are said to be \emph{mutually unbiased bases} (MUB) if for any pair of bases $\{\ket{a_k}\}_k, \{\ket{b_k}\}_k$ in the set, we have $\abs{\langle a_k | b_{k'}\rangle}^2 = \frac{1}{d} \ \forall k,k'$. 
For a set of MUB, each basis $\{\ket{\psi^i_{k}}\}_k$ induces a projective measurement $\vb{P}_i = \{\ketbra{\psi^i_{k}}\}_k$.
We refer to a POVM which is a convex combination $\vb{M} = \biguplus_i q_i \vb{P}_i$ of PMs induced by mutually unbiased bases as a \emph{MUB-POVM}.
It can be shown that any single-qubit MUB-POVM can be obtained by applying a fixed unitary operator $U$ to the effects of the LBCS POVMs~\cite{zhu_quantum_2012}, resulting in the POVM
\begin{equation} \begin{split}
q_X U^\dagger \vb{P}_X  U \uplus q_Y  U^\dagger \vb{P}_{Y} U \uplus q_Z U^\dagger \vb{P}_{Z} U  
\end{split} \end{equation}
where the notation $U^\dagger \vb{P}_i  U$ is understood element-wise, e.g., $U^\dagger \vb{P}_X  U =  \{U^\dagger\ketbra{+}{+}U, U^\dagger\ketbra{-}{-}U\}$.
In addition to the two degrees of freedom of the LBCS POVM, MUB POVMs thus carry three parameters that define the single-qubit unitary $U$, which we parametrize as 
\begin{equation}
\label{eqn:unitary_parametrization}
 U(\theta,\phi,\lambda)=\left(\begin{array}{c c}{\cos\left(\theta /2\right)} &{{-e^{i\lambda}\sin\left(\theta /2\right)}}\\ {{e^{i\phi}\sin\left(\theta /2\right)}}&{{e^{i(\phi+\lambda)}\cos\left(\theta /2\right)}} \end{array}\right).
\end{equation}

\subsection{General 6-outcome PM-simulable POVM}
Finally, the most general type of single-qubit PM-simulable POVM we consider is obtained similarly to the MUB POVMs, with a different unitary rotating each projective measurement, resulting in the POVM 
\begin{equation} \begin{split}
q_X U_X^\dagger \vb{P}_X  U_X \uplus q_Y  U_Y^\dagger \vb{P}_{Y} U_Y \uplus q_Z U_Z^\dagger \vb{P}_{Z} U_Z.
\end{split} \end{equation}
For the numerics presented in Sec.~\ref{sec:performance_limits_numerics}, we parametrize each unitary $U_X$, $U_Y$, and $U_Z$ as in Eq.~\eqref{eqn:unitary_parametrization}. 

\subsection{4-outcome dilation POVM}
For a single qubit, a four-outcome dilation POVM is represents a minimal IC measurement. 
In practice, it can be realized by applying a two-qubit dilation unitary $U_\text{dilation}$ to the targeted qubit in state $\rho$ and a second ancillary qubit that is prepared in a fixed state, typically $\ket{0}$. 
Then, both qubits are measured in the computational basis, and the four possible outcomes will occur with the probability $\text{Tr}[\rho M_k]$ of the four POVM effects $M_k$. 
The POVM effects are parametrized through the entries in the unitary $U_\text{dilation}$.
We follow the explicit parametrization of Ref.~\cite{garcia-perez_learning_2021}, which leads to eight real parameters that define the set $\{ M_k \}$.

\section{Frame theory}
\label{app:frame_theory}
Here, we review the fundamentals of frame theory %
and highlight the connection to the POVM formalism. 
Let $V$ be a vector space with an inner product $\inprod{\cdot}{\cdot}$. Let $F=\{\vb{f}_k\}_{k \in \K}$ be a set of vectors in $V$.
If there exist positive real numbers $A$ and $B$ such that $0 < A \leq B < \infty$ and
\begin{equation} \label{eq:frame_def}
    A \inprod{\vb{v}}{\vb{v}} \leq \sum_{k \in \K} \abs{\inprod{\vb{v}}{\vb{f}_k}}^2 \leq B \inprod{\vb{v}}{\vb{v}}\, \textrm{ for all } \vb{v} \in V \, , 
\end{equation}
then the set $F$ is called a \emph{frame}. Given a frame $F$, the tightest bounds $A$ and $B$ are called the \emph{frame bounds} of $F$.
This definition implies that a frame spans the vector space $V$
, which is a necessary condition for a set to be a frame. 
Note that this condition is also sufficient for finite sets in finite-dimensional vector spaces, but not in general \cite{casazza_introduction_2013}.

If the vectors constituting the frame are linearly independent, then the frame is a usual basis and it is referred to as \emph{minimally-complete}. Otherwise there is some redundancy and the frame is said to be \emph{overcomplete}. 
It can be seen as a generalization of the notion of a basis.

Let $F =\{\vb{f}_k\}_{k \in \K} \subset V$ be a frame. A frame  $D=\{\vb{d}_k\}_{k \in \K} \subset V$ such that
\begin{equation}
    \vb{v} = \sum_{k \in \K} \inprod{\vb{v}}{\vb{d}_k} \vb{f}_k = \sum_{k \in \K} \inprod{\vb{v}}{\vb{f}_k} \vb{d}_k \ \textrm{ for all } \vb{v} \in V
\end{equation}
is called a \emph{dual frame} to $F$. 
We see from the definition that if  $D$ is a dual frame to $F$, then $F$ is a dual frame to $D$.
In other words, duality is a reciprocity relation. A fundamental result of functional analysis is the existence of a dual frame for any frame \cite{li_general_1995}.
More precisely, if $F$ is minimally-complete, it has exactly one dual frame. 
If $F$ is overcomplete, it has infinitely many dual frames.  

To summarize, a frame $F$ spans $V$ and it always has a dual frame $D$. 
This offers a very convenient and natural way to find suitable coefficients to express any vector using the frame $F$. 
For any $\vb{v} \in V$,
\begin{equation}
    \vb{v} = \sum_{k \in \K} c_k \vb{f}_k \, ,  \textrm{ with } c_k = \inprod{\vb{v}}{\vb{d}_k} \textrm{ for all } k \in \K.
\end{equation}

\subsection{Frame Operator} 
\label{sec:frame_superop} 

After establishing the existence of a dual frame $D$ for any frame $F$, let us now consider explicit constructions for it.
Given a frame $\{\vb{f}_k\}_{k \in \K} \subset V$, the linear operator
\begin{equation} \begin{split}
    \F : V & \to V \\
     \vb{v} & \mapsto \sum_{k \in \K} \inprod{\vb{v}}{\vb{f}_k} \vb{f}_k
\end{split} \end{equation}
is called the \emph{canonical frame operator}.
    The canonical frame operator $\F$ is 
    \begin{enumerate}
        \item Self-adjoint : $\inprod{\vb{v}}{\F(\vb{v})} = \sum_{k \in \K} \abs{\inprod{\vb{v}}{\vb{f}_k}}^2 =\inprod{\F(\vb{v})}{\vb{v}}$ for all $\vb{v} \in V$,
        \item Positive definite: $\inprod{\vb{v}}{\F(\vb{v})} = \sum_{k \in \K} \abs{\inprod{\vb{v}}{\vb{f}_k}}^2 \geq A \norm{\vb{v}}^2 > 0$ for all $\vb{v} \in V\backslash\{\vb{0}\}$.
    \end{enumerate}
    The canonical frame operator $\F$ is thus invertible and its inverse $\F^{-1}$ is linear, self-adjoint and positive definite.
    Consider a frame $\{\vb{f}_k\}_{k \in \K} \subset V$ and its canonical frame operator $\F$. The set  of vectors 
    \begin{equation}
        \{ \vb{d}_k \in V \mid \vb{d}_k = \F^{-1}(\vb{f}_k), \, k \in \K \}
    \end{equation}
    is called the \emph{canonical dual frame} to $\{\vb{f}_k\}_{k \in \K}$. 
To see that it is indeed a dual frame to $\{\vb{f}_k\}_{k \in \K}$, simply note that we can write any $\vb{v} \in V$ as
\begin{align}
\begin{split}
    \vb{v} &= \F  \F^{-1}\vb{v} = \sum_{k \in \K} \inprod{\F^{-1}\vb{v}}{\vb{f}_k} \vb{f}_k \\ &= \sum_{k \in \K} \inprod{\vb{v}}{\F^{-1}\vb{f}_k} \vb{f}_k = \sum_{k \in \K} \inprod{\vb{v}}{\vb{d}_k} \vb{f}_k
\end{split}
\end{align}
where we used the fact that $\F^{-1}$ is self-adjoint.
    Consider now a frame $F = \{\vb{f}_k\}_{k \in \K} \subset V$ and a set of real coefficients $\{\alpha_k\}_{k \in \K} \subset \mathbb{R}$. 
    The operator
    \begin{equation} \begin{split}
        \F_\alpha : V & \to V \\
         \vb{v} & \mapsto \sum_{k \in \K} \alpha_k \inprod{\vb{v}}{\vb{f}_k} \vb{f}_k
    \end{split} \end{equation}
    is called an \emph{$\alpha$-frame operator}.
    If the operator $\F_\alpha$ is invertible, the set 
\begin{equation}
    D = \{ \vb{d}_k \in V \mid \vb{d}_k = \alpha_k \F_\alpha^{-1}(\vb{f}_k), \, k \in \K \}
\end{equation}
is a valid dual frame to $F$.
    If all the coefficients $\{ \alpha_k \}_{k \in \K}$ are positive, then the $\alpha$-frame operator will be positive definite and hence invertible.  
    Lastly, this $\alpha$-parametrization of the dual frame is invariant under uniform scaling of the coefficients. That is, the sets of coefficients $\{ \alpha_k \}_{k \in \K}$ and $\{ C \cdot \alpha_k \}_{k \in \K}$, with $C>0$, will give the same dual frame.  

\subsection{Application to POVMs} \label{sec:app_to_povm} 

As we only consider Hilbert spaces $\HS$ of finite dimension, $d < \infty$, the set of Hermitian operators is an operator-valued vector space $V=\herm$. Together with the Hilbert-Schmidt inner product, $\inprod{O_1}{O_2} = \Tr[O_1^\dagger O_2]$,  it forms a real Hilbert space of dimension $\dop = d^2$.
Therefore, the definition of a frame can be applied to the vector space $V=\herm$. In this case, a frame is a set $\{M_k\} \subset \herm$, for which there exist $A,B \in \mathbb{R}_{>0}$ such that 
\begin{equation}
    A \norm{O}_\textrm{HS}^2 \leq \sum_{k} \abs{\Tr[OM_k]}^2 \leq B \norm{O}_\textrm{HS}^2 
\end{equation}
for all $O\in \herm$.
In the following, we will only consider finite sets of operators $\{M_k\}$ acting on a finite-dimensional space $\HS$. Hence, a set $\{M_k\}_{k=1}^n$ is a frame if and only if it spans $\herm$, and
    an $n$-outcome POVM is a frame for $\herm$ if and only if it is informationally complete.

However, note that a general frame for $\herm$ is not necessarily a POVM: indeed, it does not necessarily have positive semi-definite elements nor does it necessarily sum up to the identity.
For an IC-POVM $\vb{M} = \{M_k\}_{k=1}^n$, a dual frame $\vb{D} = \{D_k\}_{k=1}^n \subset \herm$ exits such that 
\begin{equation} 
    O = \sum_{k=1}^n \Tr[OD_k] M_k = \sum_{k=1}^n \w_k M_k 
\end{equation}
for all $O\in \herm$, where the coefficients $\w_k = \Tr[OD_k]$ are real as $O, D_k \in \herm$. 

We can relax the IC condition if we are only interested in operators $O \in \spanset{\vb{M}}$. Then, there still exists a dual frame on $\spanset{\vb{M}}$ for the following reasons.
Note that $\spanset{\vb{M}}$ is a vector space and $\vb{M}$ is trivially a frame on the vector space $\spanset{\vb{M}}$. 
Therefore, there exists a dual frame $\vb{D}$ on this vector (sub-)space. Finally, as $O \in \spanset{\vb{M}}$, we can write $O = \sum_k \Tr[O D_k] M_k= \sum_{k=1}^n \w_k M_k$ by definition of a dual frame.

\section{Double-ket notation}
\label{app:double-ket_notation}
Here, we detail the vectorized double-ket notation used throughout the main text as in Ref.~\cite{dariano_classical_2005}.
The set of Hermitian operators $\textrm{Herm}(\HS)$ is an operator-valued vector space. Hence, we can associate a vector to each Hermitian operator 
\begin{equation}
    O = \sum_{i,j = 1}^d o_{i,j} \ketbra{i}{j}  \leftrightarrow  \kket{O} = \sum_{i,j = 1}^d o_{i,j} \ket{i} \otimes \ket{j} .
\end{equation}
The definition of the double-bra naturally follows  
\begin{equation}
    \bbra{O} = \sum_{i,j = 1}^d o_{i,j}^* \bra{i} \otimes \bra{j} \, .
\end{equation}
With this convention, the Hilbert-Schmidt inner product has a natural form
\begin{equation} \begin{split}
    \Tr[A^\dagger B] = 
     \sum_{j,k = 1}^d a_{j,k}^* b_{j,k}
     = \bbrakket{A}{B}  \, .
\end{split} \end{equation}
Let $\vb{M} = \big( \kket{M_1} \, \kket{M_2} \, \cdots \, \kket{M_n}\big) \in \mathbb{C}^{\dop \times n}$ denote a frame for $\herm$.
With the double-ket notation, a dual frame $\vb{D} = \big( \kket{D_1} \, \kket{D_2} \, \cdots \, \kket{D_n}\big) \in \mathbb{C}^{\dop \times n}$ is a frame that satisfies the condition
\begin{equation} \begin{split}
   & \sum_k \kket{M_k} \bbra{D_k} = \sum_k \kket{D_k} \bbra{M_k} = \mathcal{I} \\
   \iff & \vb{M} \vb{D}^\dagger = \vb{D} \vb{M}^\dagger = \mathcal{I} \, , 
\end{split} \end{equation}
where $\mathcal{I}$ denotes the identity on $\herm$.
The canonical frame superoperator $\F \in  \mathbb{C}^{\dop \times \dop}$ is given by
\begin{equation}
    \F = \vb{M} \vb{M}^\dagger = \sum_{k=1}^n \kket{M_k}\bbra{M_k}
\end{equation}
and acts on $\kket{X} \in \mathbb{C}^{\dop}$ as $\mathcal{F} \kket{X} = \sum_{k=1}^n \kket{M_k}\bbrakket{M_k}{X}$ $= \sum_{k=1}^n \Tr[M_k X]\kket{M_k}$. Then, the elements of the canonical dual frame are given by 
\begin{equation}
    \kket{D_k} = \mathcal{F}^{-1} \kket{M_k} \, , \quad k =1,2,\dots,n
\end{equation}
or, in a denser notation,
\begin{equation}
    \vb{D} = (\vb{M} \vb{M}^\dagger)^{-1} \vb{M} \, .
\end{equation}

\section{SVD parametrization of duals}
\label{app:svd_param_duals}
A general parametrization of dual frames can be obtained by resorting to the singular values decomposition (SVD). The proof given below follows closely the derivation originally presented in Ref.~\cite{krahmer_sparsity_2013}.

Let $\vb{M} = \big( \kket{M_1} \, \kket{M_2} \, \cdots \, \kket{M_n}\big) \in \mathbb{C}^{\dop \times n}$ denote a frame with bounds $A$ and $B$. By applying the SVD, we obtain
\begin{equation}
    \vb{M} = U \Sigma_{\vb{M}} V^\dagger \, , 
\end{equation}
where $U \in \mathbb{C}^{\dop \times \dop}$,  $V \in \mathbb{C}^{n \times n}$ are unitary and 
\begin{equation}
    \Sigma_{\vb{M}} = \begin{bmatrix}
            \sigma_1 & 0 & \cdots & 0 & 0 & \cdots & 0 \\
            0 & \sigma_2 &  & 0 & 0 & \cdots & 0 \\
            \vdots &  & \ddots & \vdots & \vdots &  & \vdots \\
            0 & 0 & \cdots & \sigma_\dop & 0 & \cdots & 0 \\
    \end{bmatrix} \in \mathbb{R}^{\dop \times n} \, , 
\end{equation}
where $ \sqrt{B} = \sigma_1 \geq \sigma_2 \geq \dots \geq \sigma_\dop = \sqrt{A} > 0 $ are the singular values of $\vb{M}$ \cite{casazza_introduction_2013}.

A frame $\vb{D} = \big( \kket{D_1} \, \kket{D_2} \, \cdots \, \kket{D_n}\big) \in \mathbb{C}^{\dop \times n}$ is a dual frame to $\vb{M}$ if and only if 
\begin{equation} \label{eq:cond_dual_matrix_form} \begin{split}
    & \vb{M} \vb{D}^\dagger = \vb{D} \vb{M}^\dagger = \mathds{1} \\
    \iff &  U^\dagger \vb{M} V V^\dagger \vb{D}^\dagger U  = U^\dagger  \vb{D} V V^\dagger \vb{M}^\dagger U = \mathds{1} \\
    \iff &  \Sigma_{\vb{M}} \Lambda_{\vb{D}}^\dagger = \Lambda_{\vb{D}} \Sigma_{\vb{M}}^\dagger =  \mathds{1} 
\end{split} \end{equation}
where $\Lambda_{\vb{D}} = U^\dagger  \vb{D} V$. The solutions to \eqref{eq:cond_dual_matrix_form} are $\Lambda_{\vb{D}} =$
\begin{equation} \label{eq:solution_dual_cond}
     \begin{bmatrix}
            \frac{1}{\sigma_1} & 0 & \cdots & 0 & s_{1,1} & \cdots & s_{1,n-\dop} \\
            0 & \frac{1}{\sigma_2} &  & 0 & s_{2,1} & \cdots & s_{2,n-\dop} \\
            \vdots &  & \ddots & \vdots & \vdots &  & \vdots \\
            0 & 0 & \cdots & \frac{1}{\sigma_\dop} & s_{\dop,1} & \cdots & s_{\dop,n-\dop} \\
    \end{bmatrix} \in \mathbb{C}^{\dop \times n}
\end{equation}
with arbitrary $s_{i,j} \in \mathbb{R}$ for all $i,j$. %
Hence, all dual frames to $\vb{M}$ can be expressed as
\begin{equation}
    \vb{D} = U \Lambda_{\vb{D}} V^\dagger  \in \mathbb{C}^{\dop \times n}
\end{equation}
where $\Lambda_{\vb{D}}$ has the form \eqref{eq:solution_dual_cond} and where $U, V$ depend on the frame $\vb{M}$. 
    Note that if $\vb{M}$ is a minimally-complete frame (i.e.\ $n=\dop$), then there is no freedom in the choice of $\Lambda_{\vb{D}}$ which means that $\vb{D}$ is unique, as expected. 
    By setting $s_{i,j} = 0 , \, \forall i,j $, we obtain the canonical dual frame \cite{krahmer_sparsity_2013}.    

\end{document}